\documentclass[vecphys]{svmult}

\usepackage{graphicx}        % standard LaTeX graphics tool
\usepackage[bottom]{footmisc}% places footnotes at page bottom

\newcommand{\doce}{\rm $^{12}$CO}
\newcommand{\trece}{\rm $^{13}$CO}

\newcommand{\kms}{\mbox{km~s$^{-1}$}}

\newcommand{\ms}{\mbox{$M_{\odot}$}}

\begin{document}

\title*{Deep study of the fast bipolar outflows in pre-PNe from CO mm-wave line emission}
% Use \titlerunning{Short Title} for an abbreviated version of
% your contribution title if the original one is too long
\author{G. Quintana-Lacaci\inst{1}\and
F. Jim\'enez-Esteban\inst{1}
\and
V. Bujarrabal\inst{1}
\and
A. Castro-Carrizo\inst{2}
\and
J. Alcolea\inst{3}
}
% Use \authorrunning{Short Title} for an abbreviated version of
% your contribution title if the original one is too long
\institute{Observatorio Astron\'omico Nacional (IGN), Apdo. 112, E-28803 Alcal\'a de Henares, Spain
\texttt{(g.quintana,f.jimenez-esteban,v.bujarrabal)@oan.es}
\and Institut de RadioAstronomie Millim\'etrique, 300 rue de la Piscine, 38406 Saint Martin d'H\`eres, France \texttt{ccarrizo@iram.fr}
\and Observatorio Astron\'omico Nacional (IGN), Alfonso XII N$^{\b o}$3, E-28010 Madrid, Spain
\texttt{j.alcolea@oan.es}
}
%
% Use the package "url.sty" to avoid
% problems with special characters
% used in your e-mail or web address
%
\titlerunning{Bipolar outflows in pre-PNe}
\authorrunning{Quintana-Lacaci et al.}

\maketitle

\begin{abstract}
% Your abstract goes here. Separate text sections with the standard \LaTeX\
% sectioning commands.
High spatial resolution images of PNe have shown their extremely
complex morphology. However, the circumstellar envelopes of their
progenitors, the AGB stars, are strikingly spherical. In order to
understand the carving processes leading to axisymmetric nebulae, we
are carrying out a study of a large sample of pre-PNe. Our emission
model of the nebular molecular gas ($^{12}$CO \& $^{13}$CO) will allow us to
determine important physical parameters (mass, linear
momentum, kinetic energy) of the fast bipolar and slow spherical
nebular components separately. We will study in an innovative way the
% fast bipolar outflow 
properties for each source individually, and put our results in an
evolutionary context with the help of the data obtained by us and collected from the
literature.

\end{abstract}

\section{Introduction}
\label{sec:1}
During the latter stage of stellar evolution of low- and
intermediate-mass stars, an striking transformation occurs in the
nebular morphology and kinematics: the spherical, slowly expanding
($\sim$\,15\,km\,s$^{-1}$) circumstellar envelopes of AGB stars
transform into highly aspherical PNe with fast outflows
($\sim$\,100\,km\,s$^{-1}$). High spatial resolution images
of pre-PNe has shown that the asphericity is already present in practically all
well-resolved sources \cite{ref3}. In addition, the presence of fast
bipolar outflows is a common feature in these objects
\cite{ref1}. The physical processes leading to the appearance of
this fast wind are not clearly understood, and the possible link to
other structural features such as disk or tori, or a companion star,
remain unknown. Magnetic fields may also be involved in the formation of
these outflows (see \cite{thesis}).

% In agreement with this picture, pre-PNe usually show a slow shell, probably the fossil AGB wind
% (hereafter ``spherical component''), plus a fast bipolar outflow
% (hereafter ``bipolar component'').

CO emission is the best probe to study the kinematics of the
nebular molecular gas, which is the majority of the total nebular
mass. Using a synthetic model it is possible to obtain crucial
information about the physical conditions of the nebula. CO emission
lines of pre-PNe show a characteristic profile where two components
are clearly distinguished: i) the core of the line at low velocity,
mainly (but not only) corresponding to the spherical component of
the nebula, which is probably the fossil AGB wind, and, ii) the wings 
of the line at high velocity,
corresponding to the bipolar component of the nebula. Note that
although the bipolar component is the only one detected at high
velocities, this is also detected at low velocities because of
projection effects.

The first systematic study of the physical characteristics 
% (mass, linear momentum and energy) 
of the bipolar outflows of pre-PNe from their CO emission was made by
\cite{ref1}. They considered that only the wings of the line
% , i.e. the emission at observed relative velocities larger than those
% of the spherical component, 
represent the emission of the fast bipolar outflow, and, 
%. These authors consider, moreover, 
that all the fast gas flows in the direction of the nebula axis. To go
a step further, besides increasing the number of sources in the
sample with high-quality new observations, we have modeled the CO emission 
of the bipolar component of
the nebula flowing in all directions. This way we are able to separate
the CO emission at low velocities of the bipolar component from that
of the spherical component at the core of the observed spectrum. This
will help us to obtain more accurate values of the mass, the linear
momentum, and the energy driven by the bipolar outflows, and to better
understand whether those parameters are linked to other observational
characteristics of these sources.

% Always give a unique label
% and use \ref{<label>} for cross-references
% and \cite{<label>} for bibliographic references
% use \sectionmark{}
% to alter or adjust the section heading in the running head
% Your text goes here. Use the \LaTeX\ automatism for your citations

\section{The sample}

\begin{table}[]
\caption{List of sources included in our study.}
\begin{center}
\begin{tabular}{ l  l  l  l } 
\hline
\noalign{\smallskip}
  IRAS\,Z02229+6208 & IRAS\,22223+4327 & M\,2--56        &IRAS\,19475+31 \\
  IRAS\,06530--0213 & OH\,17.7--2.0    & OH\,231.8+4.2   & IRAS\,22272+5435\\
  IRAS\,07131--0147 & IRAS\,04296+3429 & Hen\,3-401      & IRAS\,22574+6609\\
  IRAS\,08005--2356 & CRL\,618         & Roberts 22      & IRAS\,20000+3239\\
  IRAS\,18059--3211 & Frosty Leo       & HD\,101584      & IRAS\,23304+6147\\
  IRAS\,18095+2704  & IRAS\,17436+5003 & Boomerang Nebula& R\,Sct\\
  IRAS\,17438+5003  & He\,3-1475       & He\,2-113       & IRAS\,20028+3910\\
  NGC\,6302         & IRAS\,19500--1709& Mz-3            & IRAS\,23321+2545\\
  89 Her            & CRL\,2477        & M\,2--9         & IRAS\,17150--3224\\
  IRAS\,07134+1005  & CRL\,2688        & CPD -568032     & IRAS\,21282+5050\\
  Red Rectangle     & NGC\,7027        & M\,1--92        & \\
\noalign{\smallskip}
\hline

\end{tabular}
\end{center}
\end{table}

We have a sample of 43 sources (see Table\,1), most of them observed in \doce\ and \trece, 
both in transitions $J$=1--0 and $J$=2--1, at Pico de Veleta (Spain). The data corresponding 
to the rest of the sources are taken from the literature. 
% These pre-PNe are listed in Table\,1. 
With this large sample of sources it is possible to improve the statistical study on the characteristics found for their different nebular components.

\section{Model}

Starting with a few assumptions we have modeled the expected \doce\ and \trece\ 
emission profiles from the bipolar component of pre-PNe. These
assumptions were:

\begin{itemize}
\item $\vec V_{\rm exp} \propto \vec r$, i.e., Hubble expansion velocity law
% \item Mass conservation
\item Axial symmetry, ellipsoidal or bilobal
\item Optically thin emission
\end{itemize}

\begin{figure}
\centering
% Use the relevant command for your figure-insertion program
% to insert the figure file.
% For example, with the option graphics use
\includegraphics[height=4cm]{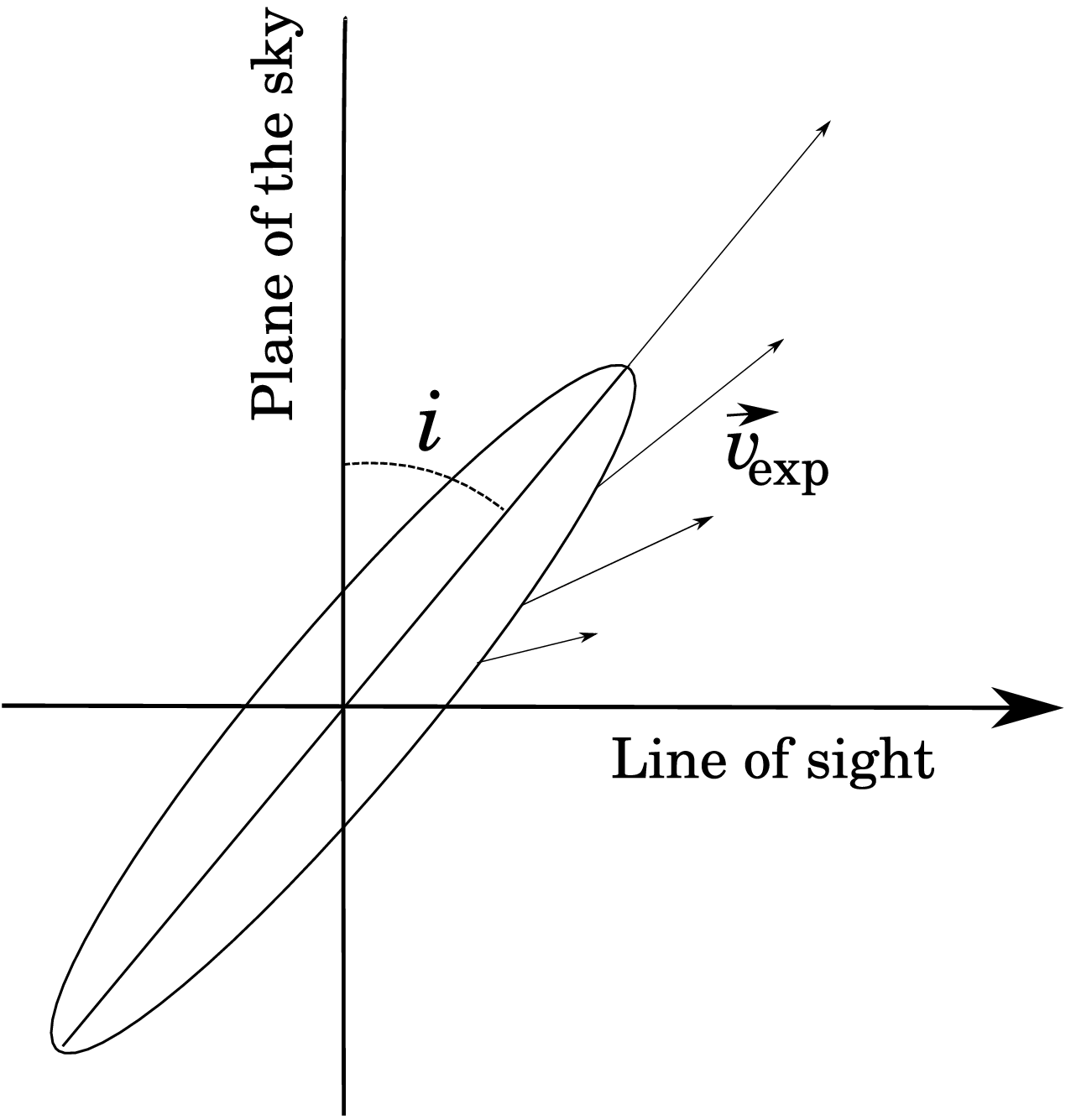}\includegraphics[height=5cm]{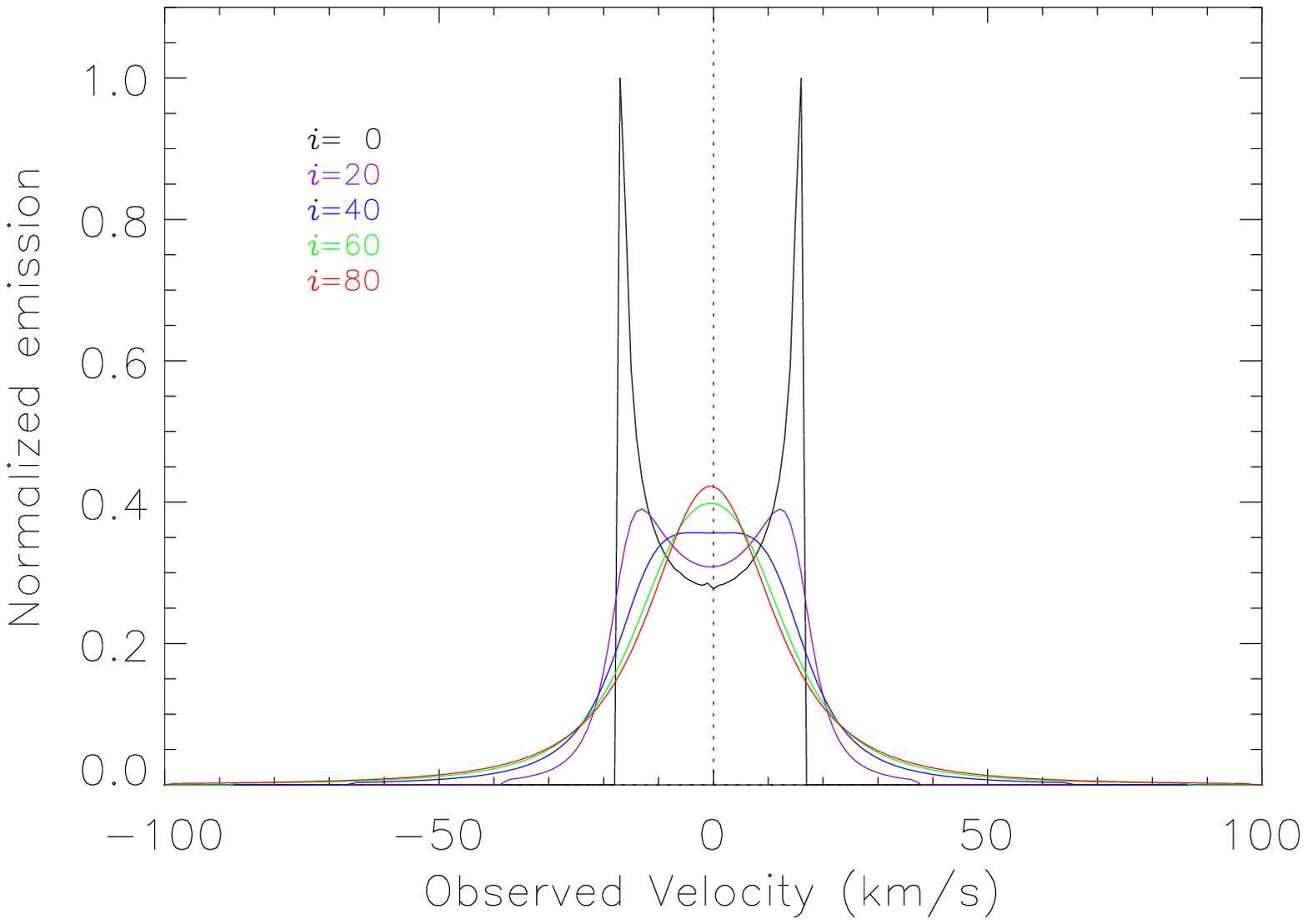}

\includegraphics[height=4cm]{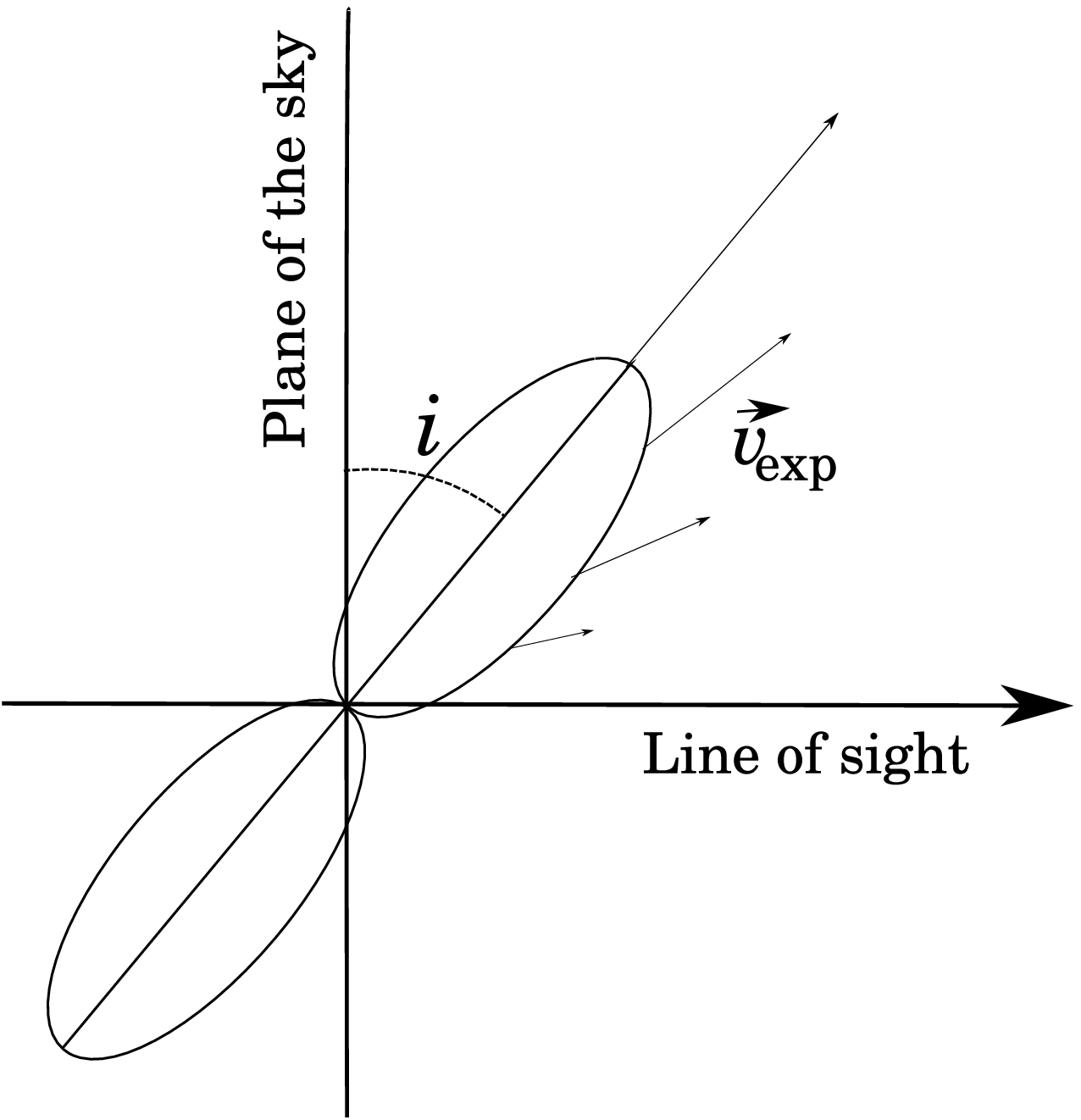}\includegraphics[height=5cm]{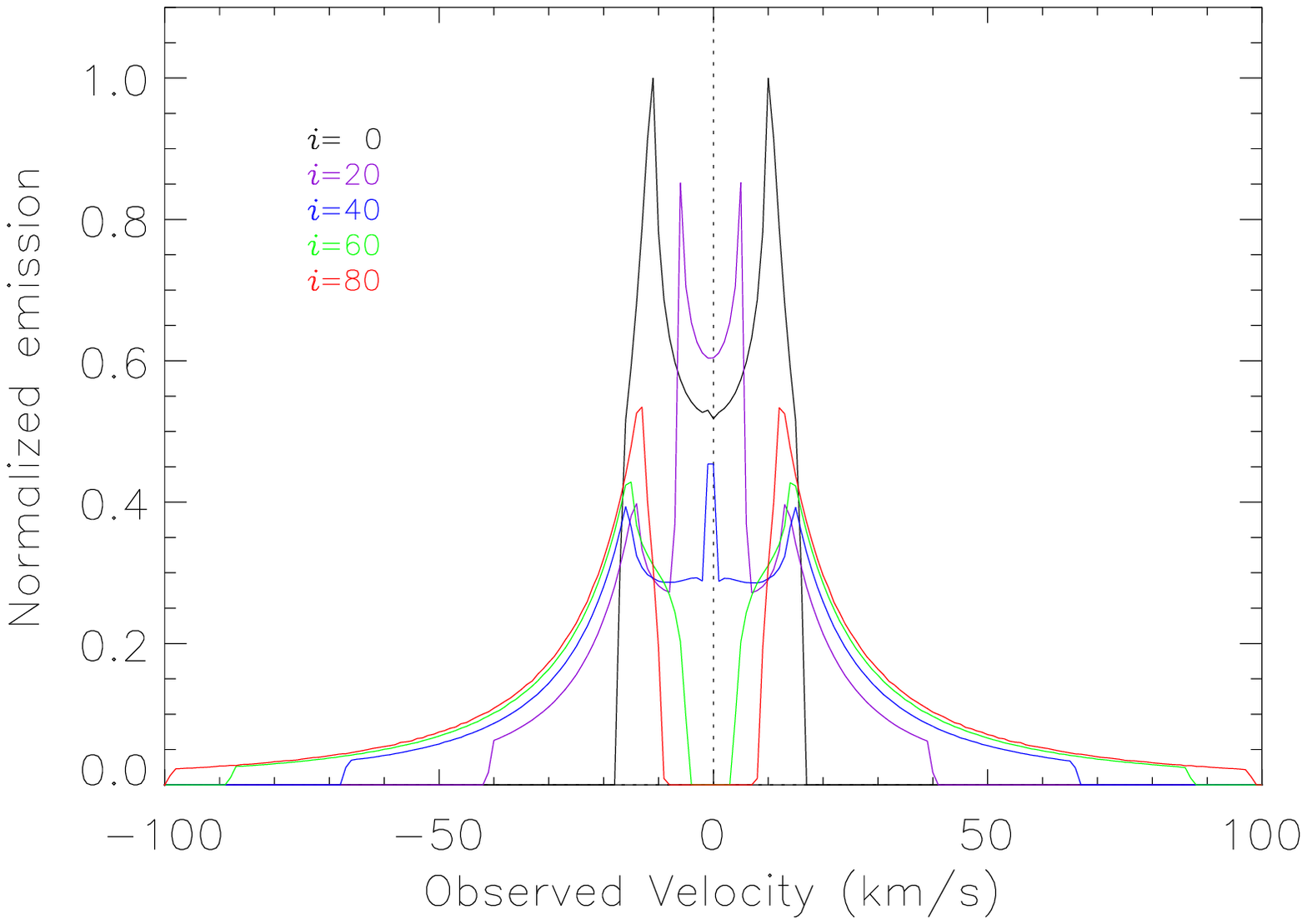}
%4
% If not, use
%\picplace{5cm}{2cm} % Give the correct figure height and width in cm
%
\caption{{\it Left:} Geometry scheme. {\it Right:} variation of the expected CO
emission profiles from the bipolar component with the inclination of
the nebula with respect to the plane of the sky.
{\it Up:} Ellipsoidal geometry. {\it Down:} Bilobal geometry}
\label{fig:1}       % Give a unique label
\end{figure}

We consider that the CO gas is confined to an infinitesimally thin
shell. Assuming mass conservation and an originally isotropical distribution of mass, 
we have a scenario similar to an anisotropically
expanding balloon, where, as it expands the density at each point
decreases following $\rho \propto r^{-2}$.

In our model, the CO gas located in the region closer to the star has
low $V_{\rm exp}$, that would correspond to the slow spherical
component. Thus, in order to avoid such CO gas when computing the fast
bipolar component emission profile, we have introduced a lower limit
to $V_{\rm exp}$, $V_{\rm AGB}$.

\section{Application of the model. The case of CRL\,2688}

\begin{figure}
\centering
\includegraphics[height=5.9cm]{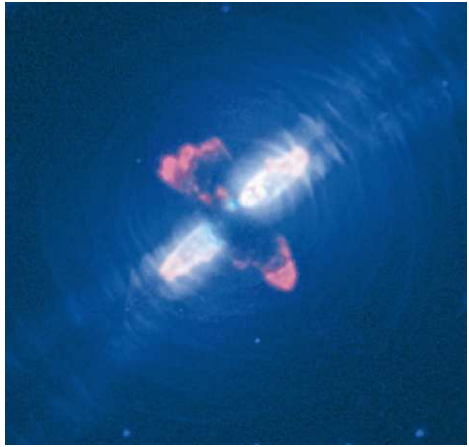}
\caption{Near-infrared image of CRL\,2688 from \cite{ref2}.}
\label{fig:2}       % Give a unique label
\end{figure}

In order to test our model, we have applied it to a well studied
source included in our sample, CRL\,2688. First we obtain from the 
bibliography the values of some input parameters of our model: i)
%, like 
the inclination
%, $i$, 
with respect to the plane of the sky, ii) the ratio between the mayor an the
minor axis of the
nebula, and iii) the distance to the source. We estimate the value of
$V_{\rm AGB}$ from the width of the core of the observed emission
line, and the rotational temperature
%, $T_{\rm rot}$, 
from the observed line intensity ratio. Once all these data are
obtained, we fit by a $\chi^2$ method our synthetic profile of the
bipolar component to the wings of the observed profile. This way we
obtain the only two free parameters: the mass and the maximum
expansion velocity of the bipolar component. After a successfully fit
is obtained, we are able to obtain the mass of the spherical component
of the nebula just by subtracting the bipolar component mass to the
total mass. The model is preferably fitted, in the cases in which it
is possible, to \trece\, lines, since they are optically thinner than
those of \doce. A factor can be introduced as a correction due to
opacity and calibration effects.

\begin{figure}
\centering

\includegraphics[angle=-90,width=11.8cm]{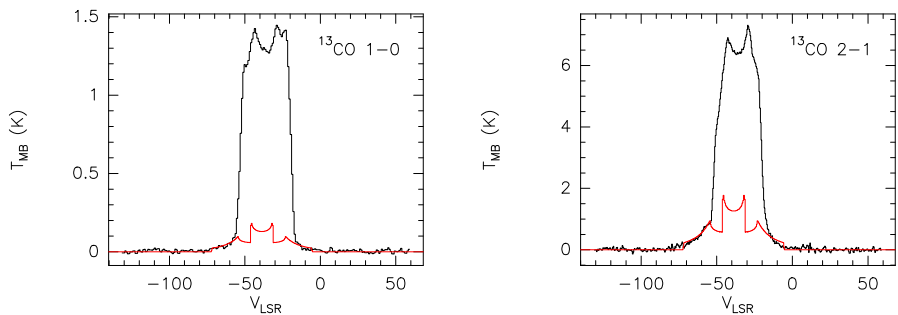}
\caption{Results of the model fitting for the wings of the profiles of
CRL\,2688. We have obtained, assuming bilobal geometry, a maximum
expansion velocity of 100\,\kms\ and a mass of 0.11\ms\ for the bipolar
component. Note the good correlation between the synthetic profile of
the bipolar component and the two observed peaks in the core of the
line.}
\label{fig:3}       % Give a unique label
\end{figure}

The results obtained for CRL\,2688 are compatible with those obtained by
\cite{ref1}. For this source they deduce a mass for the bipolar
outflows of 0.06\ms, while we obtain 0.11\ms. The main reason for this
difference is that we have not taken into account only the wings of
the spectra as the emission of the bipolar component, but also the
important contribution of the emission at low projected velocities,
which is hidden by the most intense emission of the spherical component at
the core of the line.

\section{Future work}

% As already said, 
We will use this innovative model and new observational data
to obtain the main physical
characteristics of the fast bipolar outflows of pre-PNe. Our large
sample will help us to find possible relations between them and other
observational characteristics already known for these source, and
therefore, to obtain a better understanding of how the fast collimated
bipolar outflows are formed.

%%%%%%%%%%%%%%%%%%%%%%%%%%%%%%%%%%%%%%%%%%%%%%%%%%%%%%%%%%%%%%%%%%%%%%  }

%%%%%%%%%%%%%%%%%%%%%%%%%%%%%%%%%%%%%%%%%%%%%%%%%%%%%%%%%%%%%%%%%%%%%%

\end{document}